\documentclass[aps,showpacs,preprint,footinbib,preprintnumbers]{revtex4}
\usepackage{amssymb}
\usepackage{amsmath}
\usepackage{amsfonts}
\usepackage{bm}
\usepackage{here}
\usepackage{braket}
\usepackage{fancybox} 
\usepackage{ascmac} 
\usepackage{graphicx}
\usepackage{ulem}
\usepackage[usenames]{color} 

\newcommand{\be}{\begin{equation}}
\newcommand{\ee}{\end{equation}}
\newcommand{\bea}{\begin{eqnarray}}
\newcommand{\eea}{\end{eqnarray}}
\newcommand{\del}{\partial}

\newcommand{\nonum}{\nonumber}

\begin{document}

\title{$\Lambda \left( 1405 \right)$ as a $\bar{K}N$ Feshbach resonance in the Skyrme model}

\author{Takashi Ezoe$^{1}$ and Atsushi Hosaka$^{1,2}$}
\affiliation{$^{1}$Research Center for Nuclear Physics, Osaka University, Ibaraki, 567-0048, Japan}
\affiliation{$^{2}$Advanced Science Research Center, Japan Atomic Energy Agency, Tokai, Ibaraki, 319-1195 Japan}

\date{\today}
\begin{abstract}
We describe the $\Lambda(1405)$ hyperon as a Feshbach 
resonance of a $\bar KN$ quasi-bound state coupled by a decaying channel of $\pi \Sigma$  in the Skyrme model.
A weakly bound $\bar KN$ state is generated in the laboratory frame, 
while the $\Sigma$ hyperon as a strongly bound state of $\bar KN$ 
in the intrinsic frame.  
We obtain a coupling of $\bar KN$ and $\pi \Sigma$ channels  by computing 
a baryon matrix element of the axial current.  
This coupling enables the decay of the $\bar KN$ bound state 
to $\pi \Sigma$.
It is shown that the Skyrme model supports the $\Lambda(1405)$ as a narrow Feshbach resonance.  
\end{abstract}
\maketitle
\section{Introduction}
The negative parity state of the hyperon of the lowest mass, $\Lambda(1405)$, has brought many discussions over half century, 
because its properties are not easily explained by the standard quark model~\cite{Isgur:1978xj}.  
For example, its excitation energy of about 300 MeV above the ground state $\Lambda$ hyperon with mass 1116 MeV is 
considerably smaller than the other light flavored baryons of typical excitation energy about 600 MeV, i.e., 
$N(1535) - N(940) \sim 600$ MeV.  
In fact,  before the quark model becomes popular, Dalitz and Tuan analyzed the anti-kaon and nucleon ($\bar K N$) scattering data 
and suggested the existence of a qusi-bound state of $\bar K N$ corresponding to $\Lambda(1405)$~\cite{Dalitz:1959dn,Dalitz:1960du}.  
To support such a bound state, the interaction between $\bar K$ and $N$ must be sufficiently attractive.  

Employing the mass of $\Lambda(1405)$  at  the nominal value of 1405 MeV, a $\bar K N$ potential 
was proposed to reproduce the mass in Refs~\cite{Yamazaki:2002uh,Akaishi:2002bg}
and applied to few-body systems of $\bar K$ and a few nucleons, resulting in unexpectedly deeply bound states.  
On the other hand, a chiral model for the $\bar K N$ was developed, which predicted a less attractive 
interaction that is still sufficient to generate a loosely bound $\bar K N$ state with 
a mass spectrum of $\Lambda(1405)$ being consistent with experimental data~\cite{Oset:1997it}.
In contrast with the former approach, the chiral model does not generate deeply bound strange nuclei.  
Moreover, a unique feature of the chiral models is that it generates two pole structure for 
$\Lambda(1405)$, 
one is of $\bar K N$ origin while the other $\pi\Sigma$ origin~\cite{Jido:2003cb,Hyodo:2007jq}.  
The one of  $\pi\Sigma$ origin locates at a deep imaginary region on the complex energy plane, 
resulting in a broad background structure in the spectrum.  

These different natures originate from the uncertainties in the basic interaction.  
The phenomenological interaction is determined by the nominal mass of the $\Lambda(1405)$.  
The structure of the interaction such as the ranges and strengths depend much on the data employed.  
The chiral model that is based on spontaneous breaking of chiral symmetry of QCD still
contains parameters for renormalization or subtraction.  
In both methods, parameters are adjusted to reproduce the existing data for the $\Lambda(1405)$.  

Observing this situation, we have developed an alternative approach in the Skyrme model~\cite{Ezoe:2016mkp,Ezoe:2017dnp}.
It is a non-linear field theory with chiral symmetry for mesons, where baryons emerge 
as solitons~\cite{Skyrme:1958vn,Skyrme:1961vq,Perring:1962vs,Skyrme:1962vh,Witten:1983tw,Witten:1983tx,Adkins:1983ya,Zahed:1986qz}.  
The model has been shown to be successful, at least qualitatively, for 
meson and baryon spectroscopy and their interactions. 
The advantage of this model is that once the two parameters are fixed from meson properties, 
the dynamics of baryons are determined without additional parameters.  
In this manner we expect that we better discuss exotic phenomena such as high density matter 
with knowing the origin of the dynamics.  
This is the reason that we employ the Skyrme model in the present study.  

In our previous publications~\cite{Ezoe:2016mkp,Ezoe:2017dnp}, we have investigated the $\bar K N$ interaction 
in the Skyrme model using an analogous method to the bound state approach 
by Callan and Klebanov~\cite{Callan:1985hy,Callan:1987xt}.  
Their method is formulated following the $1/N_c$ expansion 
with the collective quantization of solitons
and was shown to be successful for the descriptions of  
the ground state $\Lambda$ and $\Sigma$ hyperons.  
An interesting observation is that the $\bar K$  is strongly bound to the hedgehog soliton in its rest frame (intrinsic frame), and 
consequently $\bar K$  is interpreted as a strange quark with spin 1/2 when quantized.  
Their method corresponds in many-body physics to the  projection after variation, or the strong coupling scheme~\cite{Ring:1980aa}.  
In our approach, observing that the $\bar K$ in $\Lambda(1405)$  is weakly bound to the nucleon, 
we have proposed an alternative method, 
that is the method of projection before variation, or the weak coupling scheme.  
Setting the two parameters at suitable values,  the pion decay constant and the Skyrme parameter, 
it has been shown that the $\bar K$ feels an attractive interaction from the nucleon and is bound 
with a binding energy of order ten MeV, which is identified with  $\Lambda(1405)$.
Another interesting feature is that when the $\bar KN$ interaction is expressed in the form of a 
local potential, it exhibits  an attractive pocket at medium distances supplemented by a repulsion 
at short distances.  
These features would influence on the properties of high density matter with kaons.  

In this paper, we introduce a coupling of $\bar KN$ to $\pi \Sigma$ to enable the $\bar KN$ bound state
to decay, and investigate whether the bound $\bar KN$ state survives as a Feshbach resonance.  
In terms of the low energy method of chiral symmetry, the one pion emission decay is computed by the baryon 
matrix element of the axial current.  
Details of technical issues in performing such a computation is developed.  

\section{Actions and ansatz}
\label{sec: outline} 

Let us start with the SU(3) Skyrme model action given by~\cite{Zahed:1986qz}
\bea
	\Gamma
		 = \int d^4x \left\{ 
		 		\displaystyle{\frac{1}{16}} F_{\pi}^2 \mathrm{tr} \left( \del_{\mu}U \del^{\mu}U^{\dag} \right)
				+ \displaystyle{\frac{1}{32e^2}} \mathrm{tr}\left[ \left( \del_{\mu}U\right)U^{\dag}, \left( \del_{\nu}U\right) U^{\dag} \right]^2 
				+ L_{SB} \right\}
				+ \Gamma_{WZ}.  
\label{eq_ Skyrme action}
\eea
The first and second terms are the original Skyrme model actions
and
the third term is the symmetry breaking term due to finite masses of the pseudo-scalar mesons,
\bea
	L_{SB} = \displaystyle{\frac{1}{48}} F_{\pi}^2 \left( m_{\pi}^2 + 2 m_K^2 \right) \mathrm{tr} \left( U + U^{\dag} -2 \right)
			+ \displaystyle{\frac{\sqrt{3}}{24}} F_{\pi}^2 \left( m_{\pi}^2 - m_K^2 \right) \mathrm{tr} \left[ \lambda_8 \left( U + U^{\dag} \right) \right].
\eea
In this paper, we treat the pion as a massless particle while the kaon as massive one.
The last term in Eq.~(\ref{eq_ Skyrme action}) is the contribution of the chiral anomaly called the Wess-Zumino-Witten action 
given by~\cite{Witten:1983tw,Witten:1983tx},
\bea
	\Gamma_{WZ} = 
			\displaystyle{\frac{i N_c}{240 \pi^2}} \int d^5 x \  \varepsilon^{\mu \nu \alpha \beta \gamma}
				\mathrm{tr} \left[ \left( U^{\dag} \del_{\mu} U \right)
							\left( U^{\dag} \del_{\nu} U \right) 
							\left( U^{\dag} \del_{\alpha} U \right) 
							\left( U^{\dag} \del_{\beta} U \right) 
							\left( U^{\dag} \del_{\gamma} U \right) \right],
\eea
with $N_c$ the number of colors, $N_c = 3$.

The $\bar KN$ system is described by employing an ansatz~\cite{Callan:1985hy}
\bea
	U(x) = \xi (x)U_{K} (x)\xi(x),
\label{ansatz: ezoe hosaka}
\eea
where $\xi(x)$ is for the pion field embedded in the upper $2 \times 2$ components, 
\bea
	\xi(x)
	=
	\begin{pmatrix}
		\sqrt{U(x)} 	& 0 \\
		0		& 1
	\end{pmatrix},
	\ \ \ \ 
	U(x) = \exp \left[ 2i \bm{\tau} \cdot \bm{\pi}(x)/F_\pi \right],
\label{eq_ Hedgehog soliton in SU(3)}
\eea
with $F_\pi \sim 186$ MeV the pion decay constant, 
and $U_{K}$ for the kaon field defined by,
\bea
	U_{K}(x) 
	= 
	\exp
	\left[
		\displaystyle{\frac{2 \sqrt{2} i}{F_{\pi}}}  
		\begin{pmatrix}
			0		& K(x) \\
			K^{\dag}(x)	& 0
		\end{pmatrix}
	\right],
	\ \ \ \ 
	K(x) 
	=
	\begin{pmatrix}
		K^{+}(x) \\
		K^{0}(x)
	\end{pmatrix}.
\label{eq_ kaon nonlinear field}
\eea

The Skyrme model describes the nucleon as solitons of the pion field.  
The model accommodates a static classical solution with a specific symmetry, 
that is called the hedgehog solution, 
\bea
U_H(\bm x) = \exp \left[ i \bm{\tau} \cdot \bm{\hat  x} F(r)) \right],
\label{eq_def_U_H}
\eea
where $F(r)$ is a soliton profile function of radius $r \equiv |\bm x|$, and 
$\bm{\hat  x} = \bm x/ | \bm x |$.  
Such a classical solution does not correspond to the physical nucleons
with spin and isospin quantum numbers.  
They are generated in the collective coordinate method, 
where the variables for spin and isospin rotations of the hedgehog solution are quantized.  
Therefore, the nucleon is regarded as a rotating hedgehog, 
\bea
U_H(\bm x)  \to A(t) U_H(\bm x) A(t)^\dagger .
\label{eq_A_UH_A}
\eea
Due to the symmetry of the hedgehog solution, rotations in spin and isospin spaces are related 
leading to the constraint of equal spin ($J$) and isospin ($I$) values, $J = I$.  

In the present study for the decay $\Lambda(1405) \to \pi \Sigma$, 
we need the kaon field that plays dual roles.  
One is for $\Lambda(1405)$ where the physical $\bar K$ of isospin 1/2 is bound to the nucleon, the rotating hedgehog in the laboratory frame.
Hence we have proposed an ansatz~\cite{Ezoe:2016mkp}, 
\bea
U_{EH}(x) = A(t) \xi_H A(t)^\dagger U_K A(t) \xi_H A(t)^\dagger ,
\label{eq_def_UEH}
\eea
which is used for the construction of the $\Lambda(1405)$.  
The other is for $\Sigma$ where the $\bar K$ is bound to the hedgehog soliton 
in its intrinsic rest frame.  
The total configuration of the hedgehog soliton with a bound $\bar K$ is then 
rotated simultaneously, 
\bea
U_{CK}(x) = A(t) \xi_H  U_K \xi_H A(t)^\dagger, 
\label{eq_def_UCK}
\eea
where the subscript $CK$ is from Callan-Klebanov~\cite{Callan:1985hy}.  
This equation can be written also 
\bea
U_{CK} = (A \xi_H  A^\dagger) (A U_K A^\dagger ) (A \xi_H A^\dagger), 
\label{eq_def_UCK_mod}
\eea
which explicitly indicates that the hedgehog and kaon are rotating in the same way by the rotation matrix $A(t)$.
In terms of the two-component iso-spinor the kaon field is rotated as 
\bea
K \to A(t) K.
\label{eq_AK}
\eea

Intuitively, the two different schemes for the $\bar K$ bound in $\Lambda(1405)$ and in $\Sigma$ 
are understood by comparing 
the time for the rotating hedgehog to turn around once, $\Delta t_H$, 
and the time of the bound kaon to go around the soliton (nucleon) once, $\Delta t_K$.  
The time $\Delta t_H$ for the nucleon of spin $J = 1/2$ is estimated if we know the angular velocity $\Omega$ 
of the rotating hedgehog for the nucleon by 
$\Delta t_H \sim 2 \pi / \Omega$.
Using the relation $J = {\cal I} \Omega  = 1/2$ and the moment 
of inertia value ${\cal I} \sim 1$ fm of the rotating hedgehog, we estimate $\Omega \sim 1/2$ fm$^{-1}$ and 
hence $\Delta t_H \sim 10$ fm.  
The time $\Delta t_K$ for $\Lambda(1405)$ is estimated by using a typical binding energy of the $\bar K$
that is of order ten MeV, 
while that for $\Sigma$ is estimated by using a typical binding energy of order hundred MeV.  
We find the relation 
\bea
\Delta t_H < \Delta t_K \sim \ {\rm a \ several \ ten \ fm}
\eea
for the $\bar K$ of $\Lambda(1405)$
(the $\bar K$ goes around more slowly than the hedgehog rotates), 
implying that the $\bar K$ is 
treated as a particle moving around the rotating hedgehog in the laboratory frame.
On the other hand, 
we find 
\bea
\Delta t_H > \Delta t_K \sim \ {\rm a \ few\ fm}
\eea
 for the $\bar K$ of $\Sigma$
(the $\bar K$ goes around faster than the hedgehog rotation), 
implying that the $\bar K$ is 
treated as a particle moving around the static hedgehog in the intrinsic (rotating) frame.

\section{Coupling to the $\pi \Sigma$ channel}
\label{subsec: coupling to the piSigma channel} 
\subsection{Definitions} 

The decay of $\Lambda \left( 1405 \right) \to \pi \Sigma$ is regarded as a baryon transition accompanied by  
one pion emission, which is described by the amplitude
\bea
	\bra{\pi \Sigma} \mathcal{L}_{int} \ket{\Lambda \left( 1405 \right)}.
\label{eq_ matrix element}
\eea
To the leading order of chiral expansion in powers of small momentum, 
the interaction Lagrangian with one pion $\mathcal{L}_{int}$ is written as 
\bea
	\mathcal{L}_{int} = \displaystyle{\frac{2}{F_{\pi}}} \del_{\mu} \pi^{a} J_{5}^{\mu,a}. 
\label{eq_current_current_Lagrangian}
\eea
The isospin axial current $J_{5}^{\mu,a}$ with the isospin index $a$ is the one with the one pion pole term subtracted 
and is computed in the Skyrme model in the present study.  
We note that it is normalized in accordance with the isospin; 
for instance, for the effective interaction with the nucleon, $J_5^{\mu, a} \to \bar \psi_N \gamma_\mu \gamma_5 (\tau^a/2) \psi_N$.  
For the transition of $\Lambda \left( 1405 \right) \to \pi \Sigma$ we need the isovector axial current in the form, 
\bea 
J_5^{\mu, a} \to \bar \psi_{\Sigma}^a \gamma_\mu \psi_{\Lambda(1405)},
\eea
where $\psi_{\Sigma}^a$ and $ \psi_{\Lambda(1405)}$ are the Dirac spinors for $\Sigma$ and $\Lambda(1405)$ 
with $a$ an isospin index for $\Sigma$.  
Here $\gamma_5$ is not needed due to the negative parity of $\Lambda(1405)$.  
The baryon matrix element computed in the Skyrme model 
is then identified with the coupling constant for the effective Lagrangian
\bea
\mathcal{L}_{\Lambda(1405) \to \pi \Sigma} = g_{\Lambda(1405) \pi \Sigma} 
\frac{2}{F_{\pi}} \del^{\mu} \pi^{a} \bar \psi_{\Sigma}^a \gamma_\mu \psi_{\Lambda(1405)}.
\label{eq_L_eff_coupling}
\eea
The coupling constant $g_{\Lambda(1405) \pi \Sigma}$ is then defined to be the matrix element
\bea
g_{\Lambda(1405) \pi \Sigma}
=
 \bra{\Sigma^0}J_{\mu}^{5,3} \ket{\Lambda \left( 1405 \right)}, 
\label{eq_ matrix element of Lagrangian} 
\eea
the estimation of which is the main purpose of the present paper.

\subsection{The axial current}
\label{subsec: Axial_current}

The axial current is derived from the action Eq.~(\ref{eq_ Skyrme action}) as the Noether's current associated with the axial transformation,
\bea
	U \rightarrow g_{A} U g_{A}, \ \ \ \ g_{A} =  e^{i \bm{\theta} \cdot \bm{\lambda} /2} ,
	\label{eq_ axial transformation}
\eea
where $\bm{\lambda} = \lambda^{a} \left( a = 1, 2, 3, \cdots, 8 \right)$ and $\bm{\theta}$ are 
the Gell-Mann matrices and SU(3) parameters, respectively.
The result is $(x = (t, \bm{x}))$
\bea
	&&
	J_5^{\mu, a} (x)
	=
	\displaystyle{\frac{i F_{\pi}^2}{16}} \mathrm{tr} \left[ \lambda^{a} \left( R^{\mu} - L^{\mu} \right) \right]
	+
	\displaystyle{\frac{i}{16 e^{2}}} \mathrm{tr} 
		\left[ 
			\lambda^{a}
			\left\{
				\left[ R^{\nu}, \left[ R_{\nu}, R^{\mu} \right] \right]
				- \left[ L^{\nu}, \left[ L_{\nu}, L^{\mu} \right] \right] 
			\right\} 
		\right]
	\nonum \\
	&& \hspace{50 pt}
	-
	\displaystyle{\frac{N_{c}}{48 \pi^{2}}} 
		\epsilon^{\mu \nu \alpha \beta}
			\mathrm{tr}
			\left[ 
				\displaystyle{\frac{\lambda^{a}}{2}}
				\left( L_{\nu} L_{\alpha} L_{\beta} + R_{\nu} R_{\alpha} R_{\beta} \right)
			\right],
\label{eq_J5_general}
\eea
where 
$
R_{\mu} =  U \del_{\mu} U^{\dag}, 
L_{\mu} = U^{\dag} \del_{\mu} U
$.
The current here is regarded as an operator acting on the quantized soliton states written in terms of the collective coordinates
of rotations,  
and on the second-quantized states of the kaon as we will see below.  

Substituting the ansatz (\ref{ansatz: ezoe hosaka}) for (\ref{eq_J5_general}), 
and expanding in powers of the kaon field $K$ up to the second order,
we find
\bea
	J_5^{\mu, a} = J_5^{\mu, a, (0)} + J_5^{\mu, a, (2)} + \mathcal{O} \left( K^{3} \right), 
\eea
where superscripts $(0)$ and $(2)$ stand for the order of the kaon field. 
For our purpose, we need the second order term $J_5^{\mu, a, (2)}$ which contains two kaon fields, $K$ and $K^{\dag}$.
Moreover, in the non-relativistic approximation that we employ for baryons, 
the time component $\mu = 0$ is dominant.
The explicit form of the relevant piece of the first term of (\ref{eq_J5_general}) is 
\bea
	J_5^{0, a}(x)
	=
	\displaystyle{\frac{i}{4}} 
	\mathrm{tr} (\xi^\dagger \tau^a \xi - \xi \tau^a \xi^\dagger) 
	(K \dot K^\dagger -\dot K K^\dagger) .
\label{eq_J50_2nd}
\eea
The computation of the second and third terms of (\ref{eq_J5_general}) is tedious, but possible and is given in Appendix A.  
As anticipated in the previous section, the dual roles of the kaon fields in (\ref{eq_J50_2nd}) are implemented by 
identifying one of $K$'s in (\ref{eq_J50_2nd}) with that for $\Lambda(1405)$ and the other one for $\Sigma$, 
when computing the matrix element 
$\bra{\Sigma^{0}} J_{\mu = 0}^{5, a = 3} \ket{\Lambda \left( 1405 \right)}$.  
Explicitly, we follow the relation 
\bea
		K \ &\to& A(t) K_{CK} \ \ \ \mathrm{for \ \Sigma}, \nonumber \\
		K^\dagger &\to& K_{EH}^\dagger \ \ \ \ \ \ \ \ \  \mathrm{for \ \Lambda(1405)}.  
	\label{eq_KEH_and_KCK}
\eea
The presence of collective coordinate $A(t)$ in the first equation is inferred from (\ref{eq_AK}) and is regarded 
as a coordinate operator.  

Following the standard method for field quantization, 
the kaon fields are expanded in terms of a complete set of wavefunctions 
with the corresponding creation or annihilation operators as their coefficients.  
The field $K^\dagger_{EH}$ is regarded as an annihilation operator for the antikaon for $\Lambda(1405)$ 
and is expanded by the wavefunctions in the laboratory frame, 
\bea
	{K}_{EH}^{\dag}(t, \bm x)  = \phi_{K^-}^\dagger (t, \bm x) a_{K^-} 
	+ \phi_{\bar K^0}^\dagger (t, \bm x) a_{\bar K^0} + \cdots ,
		\label{eq_def_KEH}
\eea
where $\phi$'s and  $a$'s are the wavefunctions and the corresponding annihilation operators, respectively.  
Here we have shown only the terms of the lowest $s$-wave for the anti-kaon that are necessary for our purpose, 
\bea
\phi_{K^-}^\dagger (t, \bm x)
&=&
		\begin{pmatrix}
			1, & 0
		\end{pmatrix}
		\displaystyle{\frac{1}{\sqrt{4 \pi}}} k^*(r) e^{+ i E_{EH} t} , 
\nonum \\
\phi_{\bar K^0}^\dagger (t, \bm x)
&=&
		\begin{pmatrix}
			0, & -1
		\end{pmatrix}
		\displaystyle{\frac{1}{\sqrt{4 \pi}}} k^*(r) e^{+ i E_{EH} t} ,
		\label{eq_def_KEH_phi}
\eea 
where $k(r)$ is the $s$-wave radial function of the antikaon bound to the nucleon 
with $E_{EH}$ being the corresponding energy including its rest mass.  
The minus sign in the 2nd component for $\phi_{\bar K^0}^\dagger$ reflects the proper 
isospin transformation of $\bar K$.  

For $\Sigma$, the kaon is bound to the hedgehog with quantum numbers of the grand spin, 
the sum of isospin and orbital angular momentum, $T = I + L$.
As discussed in Ref.~\cite{Callan:1985hy}, such a bound kaon is interpreted as a strange quark 
in $p$-wave.
Therefore, 
\bea
K_{CK}(t, \bm x)  =  \phi_{s\uparrow}(t, \bm x) a^\dagger_{s\uparrow} 
	+ \phi_{s\downarrow} (t, \bm x) a^\dagger_{s\downarrow} + \cdots , 
\label{eq_def_KCK}
\eea
with
\bea
\phi_{s\uparrow}(t, \bm x) 
&=& -\sqrt{\frac{1}{4\pi}} \bm \tau \cdot {\hat {\bm x}}
\begin{pmatrix}
1\\
0
\end{pmatrix}
 s(r) e^{-iE_{CK}t} , 
 \nonum \\
 \phi_{s\downarrow}(t, \bm x) 
&=& -\sqrt{\frac{1}{4\pi}} \bm \tau \cdot {\hat {\bm x}}
\begin{pmatrix}
0\\
-1
\end{pmatrix}
s(r) e^{-iE_{CK}t} , 
\label{eq_def_KCK_phi}
\eea
where the $p$-wave nature is in the combination 
$\bm \tau \cdot {\hat {\bm x}}$, 
$s(r)$ the corresponding radial function and $E_{CK}$ the energy.  
Once again the minus sign in the lower component of the second line of (\ref{eq_def_KCK_phi}) 
reflects properly the spin transformation rule.  
The functions $k(r)$ and $s(r)$ are obtained by solving the Klein-Gordon like eigenvalue 
equations~\cite{Callan:1985hy,Callan:1987xt,Ezoe:2016mkp,Ezoe:2017dnp}.
Their normalization needs to be treated properly to reflect the structure of the Klein-Gordon like equations, 
as shown in Appendix.  

\subsection{Baryon states}
\label{sec:States}


The isosinglet state of $\Lambda(1405)$ is formed by the two isospin 1/2 states of the nucleon and the kaon, 
\bea
\ket{\Lambda \left (1405 \right)} 
&=& \sqrt{\displaystyle{\frac{1}{2}}} \ket{p K^{-}} - \sqrt{\displaystyle{\frac{1}{2}}} \ket{n \bar{K}^{0}}
\nonum \\
&=& 
\sqrt{\displaystyle{\frac{1}{2}}} \psi^N_{p\uparrow}(A) a_{K^-}^\dagger \ket{0} 
  - \sqrt{\displaystyle{\frac{1}{2}}} \psi^N_{n\uparrow}(A) a_{\bar K^0}^\dagger \ket{0} .
\label{eq_Lambda_state}
\eea
The proton ($p$) and neutron ($n$) wavefunctions with spin up and down $\psi_{pn, \uparrow \downarrow}(A)$ 
are given by the collective coordinate $A$, 
\bea
A &= &
a_0 + i\bm{\tau} \cdot \bm{a}
=
i \pi \begin{pmatrix}
- \psi^N_{n \uparrow} &  - \psi^N_{n \downarrow}  \\
\psi^N_{p \uparrow} & \psi^N_{p \downarrow} .
\end{pmatrix}.
\eea


The $\Sigma$ state is given by a combinations of diquark like wavefunctions 
of spin and isospin 1, and of the strange quark.  
For neutral spin up $\Sigma$, 
\bea
\ket{\Sigma^0(J_{3} = 1/2)} 
&=& \sqrt{\displaystyle{\frac{2}{3}}} \ket{d \left( J_{3} = 1 \right) s_{\downarrow}} 
 - \sqrt{\displaystyle{\frac{1}{3}}} \ket{d \left( J_{3} = 0 \right) s_{\uparrow}}
\nonum \\
&=& \sqrt{\displaystyle{\frac{2}{3}}} \psi^d_{10}(A) a_{s\downarrow}^\dagger\ket{0} 
 - \sqrt{\displaystyle{\frac{1}{3}}} \psi^d_{00}(A) a_{s\uparrow}^\dagger\ket{0} ,
\label{eq_Sigma_state}
\eea
where the vector-isovector diquark wavefunctions are labeled by its spin $J_3$ and isospin $I_3$, $\psi^d_{J_3 I_3}$, 
and the relevant ones here are given, 
\bea
	\psi^d_{10}(A)
	&=&
	\displaystyle{\frac{\sqrt{3}}{\pi}}
	\left( a_{1} + i a_{2} \right) \left( a_{0} + i a_{3} \right),\\
	\psi^d_{00}(A)
	&=&
	\sqrt{\displaystyle{\frac{3}{2}}} \displaystyle{\frac{i}{\pi}}
	\left( {a_{0}}^2 - {a_{1}}^{2} - {a_{2}}^{2} + {a_{3}}^{2} \right).
\eea

\section{Calculation of the matrix element}
\label{sec: calculation of matrix element}

After establishing the axial current and the wavefunction, we demonstrate how the matrix element (\ref{eq_L_eff_coupling}) is computed.
The procedure is rather straightforward, though actual computation is quite long and tedious.
Therefore, we will show the outline briefly.
Let us consider the transition to the neutral $\Sigma$ ($a = 3$).
Replace the kaon fields as in (\ref{eq_KEH_and_KCK}), and the time derivatives 
by the eigenenergies of the relevant terms of (\ref{eq_def_KEH_phi}) and (\ref{eq_def_KCK_phi}), we find
\bea
	J_5^{0,3}(x, A)
	&=& 
	-(E_{EH} + E_{CK}) \frac{1}{4}
 \mathrm{tr} 
(\xi^\dagger \tau^{3} \xi - \xi \tau^{3} \xi^\dagger) A K_{CK} K_{EH}^\dagger, \label{eq_J50_2nd_1}
\eea
where we have indicated that the current is a function of $x$ and the collective coordinate $A$.  
Using the rotating hedgehog configuration $\xi = A\xi_H A^\dagger$ with 
\bea
\xi_H = \cos \frac{F}{2} + i \bm \tau \cdot \hat {\bm x}  \sin \frac{F}{2} ,
\eea
we obtain
\bea
J_5^{0,3}(x, A)
&=& 
-(E_{EH} + E_{CK}) \frac{\sin (F/2)}{4}
 \mathrm{tr} 
(\tau^{3} \bm{\tau}\cdot \hat {\bm x}^\prime - \bm{\tau}\cdot \hat {\bm x}^\prime\tau^{3} )
AK_{CK} K_{EH}^{\dag} ,
\label{eq_J50_2nd_2}
\eea
where 
$
\bm{\tau}\cdot \hat {\bm x}^\prime = A\bm{\tau}\cdot \hat {\bm x}A^\dagger .
$

For the transition amplitude, we need to take the matrix element of the interaction Lagrangian 
(\ref{eq_current_current_Lagrangian}) with the initial $\Lambda(1405)$ and the final $\Sigma \pi$, 
with a finite pion momentum $q^\mu = (E_\pi, \bm q$), $E_\pi = \sqrt{m_\pi^2 + \bm q^2}$.  
Performing necessary trace algebra for the relevant $2 \times 2$ matrices, 
we integrate over the space-time $d^4x$ and collective coordinates $d\mu(A)$, where 
\bea
	\int d \mu \left( A \right) = \int_{0}^{\pi} d \theta_{1}  d \theta_{2} \int_{0}^{2 \pi} d \theta_{3} \sin^{2} \theta_{1} \sin \theta_{2},
\eea
and the relation between the three angles and the SU(2) rotation matrix is given by
\bea
		a_{0} &=& \cos \theta_{1} \nonum \\
		a_{1} &=& \sin \theta_{1} \sin \theta_{2} \cos \theta_{3} \nonum \\
		a_{2} &=& \sin \theta_{1} \sin \theta_{2} \sin \theta_{3} \nonum \\
		a_{3} &=& \sin \theta_{1} \cos \theta_{2}. \nonum
\eea
The time integral leads to the $\delta$-function for energy conservation.  
After these manipulations, we arrive at a rather compact expression
\bea
&&
	\bra{\pi^{0}(\bm q) \Sigma^{0}} {\cal L}_{int} \ket{\Lambda(1405)}
	\nonum \\
& &=
	\displaystyle{\frac{2}{F_{\pi}}}
	\int d^4 x d\mu(A ) \ \bra{\pi} \del^{0} \pi^{3}(x) \ket{0} \bra{\Sigma} J_5^{0,3}(x,A) \ket{\Lambda(1405)}
	\nonum \\
& & = i\delta(E_{\pi} + E_{CK} - E_{EH} )
 \displaystyle{\frac{2}{F_{\pi}}}
	\int_{0}^{\infty} d r \ r^{2} 
j_0(qr)
	\frac{E_{\pi} \left( E_{{EH}} + E_{CK} \right) }{9}
		\sin F
		s(r) k^*(r) .
\label{eq_ME_Lint(2)}
\eea
The presence of the spherical Bessel function $j_0(qr) = \sin( q r )/(q r)$ indicates that the decaying pion is 
in the $s$-wave as it should be.  

So far we have shown the result for the second order derivative term.  
The computation goes similarly for the Skyrme and WZW terms.  
The results are summarized as follows, 
\bea
	&&
	\bra{\pi^{0}(\bm q) \Sigma^{0}} {\cal L}_{int} \ket{\Lambda(1405)}
	\nonum \\
	&=& 
	\displaystyle{\frac{2}{F_{\pi}}}
	\int d^4 x d\mu(A ) \ \bra{\pi} \del^{0} \pi^{3}(x) \ket{0} \bra{\Sigma} J_5^{0,3}(x,A)^{ (2+4+WZW)} \ket{\Lambda(1405)}
	\nonum \\
	&=& i\delta(E_{\pi} + E_{CK} - E_{EH} ) 
	\displaystyle{\frac{2}{F_{\pi}}}
	\int_{0}^{\infty} d r \ r^{2} 
j_0(qr)
\left(
{\cal I}_2 + {\cal I}_2  + {\cal I}_{WZW}
\right) , 
\label{eq_ME_Lint(total)}
\eea
where 
\bea
{\cal I}_2 &=& \frac{E_{\pi} \left( E_{{EH}} + E_{CK} \right) }{9} \sin F s(r) k^*(r) , 
\nonum \\
{\cal I}_4 &=&
		- \displaystyle{\frac{\left( E_{s} + E_{\bar{K}} \right)}{3}} s(r) k^*(r)
			\left[
				\displaystyle{\frac{2}{3}} \sin F 
					\left\{
						\left( F^{\prime} \right)^{2} + \displaystyle{\frac{\sin^{2} F}{r^{2}}}
					\right\}
			\right]
		\nonum \\
		&+& \displaystyle{\frac{E_{\bar{K}}}{3}} s(r) k^*(r)
				\left[ 
					\displaystyle{\frac{4}{3}} \displaystyle{\frac{c^{2} \sin F}{r^{2}}} \left( 5 c^{2} - s^{2} \right)
				\right]
		+ \displaystyle{\frac{E_{s}}{3}} s(r) k^*(r)
			\left[ 
				\displaystyle{\frac{4}{3}} \displaystyle{\frac{s^{2} \sin F}{r^{2}}} \left(- c^{2} + 5 s^{2} \right)
			\right]
		\nonum \\
		&-& \displaystyle{\frac{E_{\bar{K}}}{3}} s^{\prime}(r) k^*(r)
			\left[
				2 F^{\prime}
				\left(
					- \displaystyle{\frac{5 c^{2}}{3}}
					+ \displaystyle{\frac{7 s^{2}}{3}}
				\right)
			\right]
		+ \displaystyle{\frac{E_{s}}{3}} s(r) k^{* \prime}(r)
				\left[ 
					2 F^{\prime}
					\left(
						\displaystyle{\frac{7 c^{2}}{3}} - \displaystyle{\frac{5 s^{2}}{3}}
					\right)
				\right],
\nonum \\
{\cal I}_{WZW} 
&=&
\left(
s(r) k^{\ast}(r) \frac{4 \sin F F^{\prime}}{r^{2}}
+
s^{\prime}(r) k^*(r) \frac{2 \sin^{2} F}{r^{2}}
- s(r) k^{* \prime}(r)\frac{2 \sin^{2} F}{r^{2}}
\right).
\eea
Here we have introduced the notation $c = \cos(F/2)$ and  $s = \sin(F/2)$.

\section{Results and discussions}
\label{sec: Results and discussions}

In this section, we present and discuss our numerical results for the decay of $\Lambda(1405) \to \pi \Sigma$.  
The formulae that are derived in the previous sections determine the coupling constant 
$g_{\Lambda(1405) \pi \Sigma}$ as defined 
in the effective Lagrangian (\ref{eq_L_eff_coupling}).  
The decay width is then computed by the formula, 
\bea
	\Gamma_{\Lambda(1405) \rightarrow \pi \Sigma} 
	&=& 
	g_{\Lambda \left( 1405 \right)\pi \Sigma}^{2} \displaystyle{\frac{| \bm{q} |}{\pi}}
		\displaystyle{\frac{E_{\Sigma} + m_{\Sigma} }{4 \left( E_{\Sigma} + E_{\pi} \right)}} \times 3.
\eea
The factor 3 is for isospin sum.   
For kinematic parameters we employ the physical values that are fixed by the experiment as 
summarized in Table \ref{tab_kinematic_parameters}.  
Here we take the mass of $\Lambda(1405)$ slightly higher than the nominal value, that is 1420 MeV, 
considering the recent discussions of the two pole structure of $\Lambda(1405)$, and the 
$\bar K N$ quasi-bound state is considered to locate at around the higher mass region~\cite{Tanabashi:2018oca}.  
\begin{table}[H]
\begin{center}
\begin{tabular}{ c  c  c  c  c  c  c }  \hline
$m_{\pi}$	& $m_{K}$   & $m_{\Sigma}$ & $m_{\Lambda(1405)}$  & $| \bm{q} |$ & $E_{\pi}$ & $E_{\Sigma}$
\\ \hline
138	& 49 &1193 & 1420 & 166	& 216 & 1204	\\ \hline
\end{tabular}
\caption{Kinematical inputs for the decay of $\Lambda(1405)$ in units of MeV.}
\label{tab_kinematic_parameters}
\end{center}
\end{table}

Our main results in this paper are shown in Table~\ref{tab_results}, 
where various contributions to the coupling constants and the resulting decay widths are given for three 
sets of the Skyrme model parameters, A, B and C. 
In Set A, the decay constant $F_\pi$ is taken at an average of the pion and kaon decay constants, 
while in Set B it is set at the pion decay constant.
The Set C is from Ref.~\cite{Adkins:1983ya}.
In all cases, the Skyrme parameter $e$ is determined such that the $N\Delta$ mass splitting 
is reproduced.   

\begin{table}[H]
\begin{center}
\begin{tabular}{ c  c  c  c  c c c c c}  \hline
& $F_{\pi}$(MeV)  & $e$ & B.E. (MeV) & $g_2$ & $g_4$ & $g_{WZW}$ & $g_{\Lambda(1405) \pi \Sigma}$ & $\Gamma$(MeV)\\ \hline
Set A & 205  & 4.67	& 20.6  & 0.0545 \ & 0.0385 \ &0.0938  & 0.187 & 2.3 \\ 
Set B & 186  & 4.82 & 32.2  & 0.0609 \ & 0.0439 \ & 0.1180 & 0.223 & 3.3 \\ 
Set C & 129  & 5.45 & 81.3  & 0.0437 \ & 0.0520 \ &0.2371 & 0.333 & 7.4 \\ \hline
Data & 186 & 5.75 & 30 &  &  &  &  0.87 ($\sim$0.55) \ & \ 50.5 ($\sim$20) \\ \hline
\end{tabular}
\caption{Results for the three sets of  Skyrme model parameters.  
Contributions of the coupling constant from the 2nd order, 4th order and WZW terms are shown separately as $g_2$,  $g_4$ and $g_{WZW}$.
The experimental data for $\Lambda(1405)$ are taken from the averaged value from PDG and the corresponding coupling constant $g_{\Lambda(1405) \pi \Sigma}$
is evaluated by them.  
The numbers in the parentheses are those expected for the $\bar K N$ dominant pole.
The data for $e$ is also shown when identifying it with the coupling of $\rho \to \pi \pi$ decay~\cite{Hosaka:2001ux}.  
}
\label{tab_results}
\end{center}
\end{table}

As seen from the Table ~\ref{tab_results}, the present model predictions of the decay width $\Gamma$ are small as compared to 
the experimental data, and scatter in a range from the minimum value to the maximum value that is about three times larger than the minimum value.  
The experimental data is taken from PDG where they quote the average number $50.5 \pm 2.0$ MeV~\cite{Tanabashi:2018oca}.
There are, however, discussions about the two pole structure of $\Lambda(1405)$ having the 
$\bar K N$ and $\pi \Sigma$ origin.  
The $\bar K N$ originated one locates relatively higher in mass at around 1420 MeV and has a narrower width, 
while the $\pi \Sigma$ originated one locates lower with a wider width.  
Our present result is to be compared with the former $\bar K N$ dominant one, whose width is 
expected to be around 20 MeV~\cite{Tanabashi:2018oca}. 
Thus the corresponding coupling constants are shown in parentheses.  

The reason that the model predictions scatter in a rather wide range is that the amplitude is 
proportional to $1/F_\pi$ and that the overlap integral in the matrix element is  sensitive to the structure of the kaon wavefunctions 
of $\Lambda(1405)$ and of $\Sigma$.  
It is not difficult to see that these factors may change the coupling constant by a few times.  
Then a possible reasons for small values may be explained by the overlap integral; 
in the present approach the two limits are employed for the construction of the wavefunctions of $\Lambda(1405)$ 
and $\Sigma$, the weak coupling and strong coupling limits.  
The matrix elements for the transition amplitudes computed by the integral of the  two wavefunctions 
is therefore suppressed.
In realistic situation, the both wavefunctions are between the two limits and therefore 
the overlap integral would gain some  strength.  
We also consider that the suppression is related to the bound state approach where 
the kaon is regarded as a heavy meson and is, as well as hyperons, not treated as flavor SU(3) multiplets.  
Physically, the transition from $\bar KN$ to $\pi \Sigma$ requires an exchange of a (heavy) strange quark 
from $\bar K$ to $\Sigma$.
It is natural to consider that such a heavy particle exchange is suppressed.  

Aside from the quantitative aspect, it is worth emphasizing as the main conclusion of the present study 
that the resulting decay width turns out to be narrow. 
This enables for the $\bar K N$ bound state to remain as a Feshbach resonance, seemingly a natural 
consequence that the Skyrme model supports.

%


\begin{acknowledgments}
This work has been supported in part by Grants-in Aid for Scientific Research,  Grants No. 17K05441(C) and by Grants-in Aid for Scientific Research on Innovative Areas (No. 18H05407).

\end{acknowledgments}

\appendix
\section{Explicit expressions of the axial current}
\label{App: Explicit expressions of the axial current}


In this appendix, we show the explicit form of the axial current, 
\bea
	&&
	J_5^{\mu, a} 
	=
	\displaystyle{\frac{i F_{\pi}^2}{16}} \mathrm{tr} \left[ \lambda^{a} \left( R^{\mu} - L^{\mu} \right) \right]
	+
	\displaystyle{\frac{i}{16 e^{2}}} \mathrm{tr} 
		\left[ 
			\lambda^{a}
			\left\{
				\left[ R^{\nu}, \left[ R_{\nu}, R^{\mu} \right] \right]
				- \left[ L^{\nu}, \left[ L_{\nu}, L^{\mu} \right] \right] 
			\right\} 
		\right]
	\nonum \\
	&& \hspace{50 pt}
	-
	\displaystyle{\frac{N_{c}}{48 \pi^{2}}} 
		\epsilon^{\mu \nu \alpha \beta}
			\mathrm{tr}
			\left[ 
				\displaystyle{\frac{\lambda^{a}}{2}}
				\left( L_{\nu} L_{\alpha} L_{\beta} + R_{\nu} R_{\alpha} R_{\beta} \right)
			\right] .
\eea
%
The term from the second derivative term has been already given in (\ref{eq_J50_2nd}) for $\mu = 0$, 
\bea
	J_5^{0, a}({\rm 2nd})
	=
	\displaystyle{\frac{i}{4}} 
	\mathrm{tr} (\xi^\dagger \tau^a \xi - \xi \tau^a \xi^\dagger) 
	(K \dot K^\dagger -\dot K K^\dagger) .
\eea
%
%
%
For the term from the fourth derivative (Skyrme) term, we find
\bea
	J_5^{0, a}({\rm 4th})
	&=& 
	- \frac{i }{4 e^{2}F_\pi^2} \mathrm{tr} 
			\left(
				\lambda^{a}
					\left[ \alpha^{i, (0)}, \left[ \alpha_{ i}^{(0)}, \alpha^{ 0, (2)} \right] \right]
			\right.
		\nonum \\
	&+& 
			\left.
			2 \lambda^{a}
					\left[ \alpha^{ i, (0)}, \left[ \alpha_{ i}^{(1)}, \alpha^{ 0, (1)} \right] \right]
			+ 2 \lambda^{a}\left[ \alpha^{ i, (1)}, \left[ \alpha_{ i}^{(0)}, \alpha^{ 0, (1)} \right] \right]
			\right) - \left( \xi \leftrightarrow \xi^{\dag}\right),
\eea
where
\bea
	\alpha_{ i}^{(0)}
		&=&
			\begin{pmatrix}
				\tilde{U}_{H} \del_{i} \tilde{U}_{H}^{\dag}	& 0\\
				0 								& 0
			\end{pmatrix} , 
	\\
	\alpha_{ i}^{(1)}
		&=&
			\begin{pmatrix}
				0																				& - \tilde{U}_{H} \del_{i} \left( \tilde{\xi}^{\dag} A K_{CK} \right) \\
				 K_{EH}^{\dag} \tilde{\xi} \del_{i} \tilde{U}_{H}^{\dag} - \del_{i} \left( K_{EH}^{\dag} \tilde{\xi}^{\dag} \right)	& 0
			\end{pmatrix} , 
	\\
	\alpha_{0}^{(1)}
		&=&
			\begin{pmatrix}
				0								& - \tilde{\xi} A \del_{0} \left( K_{CK} \right) \\
				-  \del_{0} K_{EH}^{\dag} \tilde{\xi}^{\dag}	& 0
			\end{pmatrix} , 
	\\
	\alpha_{0}^{(2)}
		&=& 
			\begin{pmatrix}
			 \tilde{\xi} A\dot K_{CK} K_{EH}^{\dag} \tilde{\xi}^{\dag} - \tilde{\xi} A K_{CK}\dot K_{EH}^{\dag} \tilde{\xi}^{\dag}	
			 					& 0 \\
			0 																					
								& - K_{EH}^{\dag} A\dot K_{CK} + \dot K_{EH}^{\dag}  A K_{CK} 
			\end{pmatrix} .
\eea
For the term from the WZW term, we find
\bea
	J^{0,a}_{5}({\rm WZW})
		&=&
		\frac{N_{c}\epsilon^{i j k}}{24 \pi^{2} F_\pi^2}
		\left[ \lambda^{a}
		\left(\beta_{i}^{(0)} \beta_{j}^{(0)} \beta_{k}^{(2)} 
		+ \beta_{i}^{(0)} \beta_{j}^{(2)} \beta_{k}^{(0)} 
		+ \beta_{i}^{(2)} \beta_{j}^{(0)} \beta_{k}^{(0)} \right)
		\right. \nonum \\
		&& +
		\left. 2 \lambda^{a}
		\left( \beta_{i}^{(0)} \beta_{j}^{(1)} \beta_{k}^{(1)} 
		+ \beta_{i}^{(1)} \beta_{j}^{(0)} \beta_{k}^{(1)} 
		+ \beta_{i}^{(1)} \beta_{j}^{(1)} \beta_{k}^{(0)} \right)
		\right] + \left\{ \xi \leftrightarrow \xi^{\dag} \right\},
\label{eq_ axial current from WZ}
\eea
where
\bea
	\beta_{\mu}^{(0)} 
		&=& U_{\pi}^{\dag} \del_{\mu}  U_{\pi} 
			=
			\begin{pmatrix}
				L_{\mu} 	& 0\\
				0 		& 0
			\end{pmatrix} , 
		\\
	\beta_{\mu}^{(1)} 
		&=& 
			\begin{pmatrix}
				0														& - {\xi^{\dag}}^{2} \del_{\mu} \left( \xi K \right) \\
				K^{\dag} \xi^{\dag} \del_{\mu} \xi^{2} -\del_{\mu} \left( K^{\dag} \xi \right)	& 0
			\end{pmatrix} , 
		\\
	\beta_{\mu}^{(2)}
		&=& 
			\begin{pmatrix}
			 {\xi^{\dag}}^{2} \del_{\mu} \left( \xi K \right) K^{\dag} \xi - \xi^{\dag} K \del_{\mu} \left( K^{\dag} \xi^{\dag} \right) {\xi}^{2}	
			 	& 0 \\
			0 	
				& - 2 K^{\dag} \xi^{\dag} \del_{\mu} \left( \xi K \right) + \del_{\mu} \left( K^{\dag} K \right)
			\end{pmatrix} .
\eea

\section{Normalization conditions}
\label{app: Normalization conditions}

In this appendix, we show the normalization conditions for the kaon and $s$-quark wavefunctions 
which are consistent with the solutions of the Klein-Gordon equation, and with the canonical commutation relations.
First, in the CK approach, the normalization is given by~\cite{Callan:1985hy,Callan:1987xt},
\bea
		4 \pi \int dr r^2 s_{n}^{\ast} (\bm r) s_{m} (\bm r)
			\left[ f (r) \left( \omega_{n} + \omega_{m} \right) + 2 \lambda (r) \right] 
			&=& \delta_{nm},  \nonum \\
		4 \pi \int dr r^2 \tilde{s}_{n}^{\ast} (\bm r) \tilde{s}_{m}(\bm r)
			 \left[ f (r) \left( \tilde{\omega}_{n} + \tilde{\omega}_{m} \right) - 2 \lambda(r) \right] 
			&=& \delta_{nm},  \\
		4 \pi \int dr r^2 s_{n}^{\ast}(\bm r) \tilde{s}_{m} (\bm r)
			\left[ f(r) \left( \omega_{n} - \tilde{\omega}_{m} \right) + 2 \lambda(r) \right] 
			&=& 0,\nonum
\label{eq_ orthonormal condition in the CK}
\eea
where $s_{m} (\bm{r})$ and $\tilde{s}_{m}(\bm{r})$  are the wavefunctions of $s$ and $\bar{s}$-quark in the $m$-mode, respectively, 
and  $\omega_{m}$ and $\tilde{\omega}_{m}$ the corresponding wigenenergies, 
The radial dependent functions $f(r)$ and $\lambda(r)$ are given by
\bea
	f \left( r \right) 
		&=& 1 + \displaystyle{\frac{1}{\left( e F_{\pi} \right)^{2}}} \left[ 2 \displaystyle{\frac{\sin^{2} F}{r^{2}}} + F'^{2} \right],
	\\
	\lambda \left( r \right)
		&=& - \displaystyle{\frac{N_{c} E}{2 \pi^{2} {F_{\pi}}^{2}}}  \displaystyle{\frac{\sin^{2} F}{r^{2}}} F' .
\eea

In the $EH$ approach, the normalization conditions are given by,
\bea
		4 \pi \int dr r^2 k_{n}^{\ast}(\bm r) k_{m}(\bm r)
			\left[ 
				f \left( \omega_{n} + \omega_{m} \right) 
				+ 2 \left\{ \rho_{1} + \lambda_{1} \right\} 
				- \displaystyle{\frac{1}{r^{2}}} \displaystyle{\frac{d}{dr}} \left( r^{2} \rho_{2} \right)
			\right] &=& \delta_{nm} , 
		\\
		4 \pi \int dr r^2 \tilde{k}_{n}^{\ast}(\bm r) \tilde{k}_{m}(\bm r)
			\left[ 
				f \left( \tilde{\omega}_{n} + \tilde{\omega}_{m} \right) 
				- 2 \left\{ \rho_{1} + \lambda_{1} \right\} 
				+ \displaystyle{\frac{1}{r^{2}}} \displaystyle{\frac{d}{dr}} \left( r^{2} \rho_{2} \right)
			\right] &=& \delta_{nm} ,
		\\
		4 \pi \int dr r^2 k_{n}^{\ast} (\bm r) \tilde{k}_{m}(\bm r)
			\left[ 
				f \left( \omega_{n} - \tilde{\omega}_{m} \right) 
				+ 2 \left\{ \rho_{1} + \lambda_{1} \right\} 
				- \displaystyle{\frac{1}{r^{2}}} \displaystyle{\frac{d}{dr}} \left( r^{2} \rho_{2} \right)
			\right] &=& 0, \nonum
\label{eq_ orthonormal condition in the EH}
\eea
where
\bea
	\rho_{1}(r)
		&=& - \displaystyle{\frac{4 \sin^{2} \left( F / 2 \right)}{3 \Lambda}} \bm{I}_{K} \cdot \bm{I}_{N}
				\left[
					1 + \displaystyle{\frac{1}{\left( e F_{\pi} \right)^{2}}}
							\left( 
								\displaystyle{\frac{4}{r^{2}}} \sin^{2} F + {F^{\prime}}^{2}
							\right)
				\right]
			\nonum \\
			&& \hspace{30 pt}
			- \displaystyle{\frac{\sin^{2} \left( F /2 \right)}{\Lambda}}
				\left[
					1 + \displaystyle{\frac{1}{\left( e F_{\pi} \right)^{2}}}  
							\left( 
								\displaystyle{\frac{5}{r^{2}}} \sin^{2} F + {F^{\prime}}^{2}
							\right)
				\right] , 
	\\
	\rho_{2}(r)
		&=& \displaystyle{\frac{1}{\left( e F_{\pi} \right)^{2}}} 
			\left[ 
				\displaystyle{\frac{\sin F}{\Lambda}} F^{\prime} 
					\left(
						4 \bm{I}_{K} \cdot \bm{I}_{N} + 3 
					\right)
			\right] , 
	\\
	\lambda_{1}(r)
		&=& \displaystyle{\frac{N_{c}}{{F_{\pi}}^{2}}} B^{0}, \ \ \ 
		B^{0} =  - \displaystyle{\frac{1}{2\pi^{2}}}  \displaystyle{\frac{\sin^{2} F}{r^{2}}} F' , 
\eea
where $k_{m}(\bm{r})$  and $\omega_{m}$ are the wavefunctions and the corresponding eigenenergies, respectively, 
and the tilded variables are for the kaon. 

These normalization conditions Eqs.~(\ref{eq_ orthonormal condition in the CK}) and~(\ref{eq_ orthonormal condition in the EH}) is obtained 
in order to satisfied with the canonical quantization condition,
\bea
	\left[ k_{n} (\bm{r}, t ), \pi_{m}(\bm{r}^{\prime}, t ) \right] = i \delta_{nm} \delta^{\left( 3 \right)}(\bm{r} - \bm{r}^{\prime}),
\eea
where $\pi_{m}(\bm{r}^{\prime}, t)$ is the canonical momentum conjugate to $k_{m}(\bm{r}, t)$.



\begin{thebibliography}{99}
\bibitem{Isgur:1978xj} 
N.~Isgur and G.~Karl,
Phys.\ Rev.\ D {\bf 18}, 4187 (1978).
doi:10.1103/PhysRevD.18.4187

\bibitem{Dalitz:1959dn} 
R.~H.~Dalitz and S.~F.~Tuan,
Phys.\ Rev.\ Lett.\  {\bf 2}, 425 (1959).
doi:10.1103/PhysRevLett.2.425

\bibitem{Dalitz:1960du} 
R.~H.~Dalitz and S.~F.~Tuan,
Annals Phys.\  {\bf 10}, 307 (1960).
doi:10.1016/0003-4916(60)90001-4
 
\bibitem{Yamazaki:2002uh} 
T.~Yamazaki and Y.~Akaishi,
Phys.\ Lett.\ B {\bf 535}, 70 (2002).
doi:10.1016/S0370-2693(02)01738-0

\bibitem{Akaishi:2002bg} 
Y.~Akaishi and T.~Yamazaki,
Phys.\ Rev.\ C {\bf 65}, 044005 (2002).
doi:10.1103/PhysRevC.65.044005

\bibitem{Oset:1997it}
E.~Oset and A.~Ramos,
Nucl. Phys. A \textbf{635}, 99-120 (1998)
doi:10.1016/S0375-9474(98)00170-5
[arXiv:nucl-th/9711022 [nucl-th]].

\bibitem{Jido:2003cb}
D.~Jido, J.~Oller, E.~Oset, A.~Ramos and U.~Meissner,
Nucl. Phys. A \textbf{725}, 181-200 (2003)
doi:10.1016/S0375-9474(03)01598-7
[arXiv:nucl-th/0303062 [nucl-th]].

\bibitem{Hyodo:2007jq}
T.~Hyodo and W.~Weise,
Phys. Rev. C \textbf{77}, 035204 (2008)
doi:10.1103/PhysRevC.77.035204
[arXiv:0712.1613 [nucl-th]].

\bibitem{Tanabashi:2018oca}
M.~Tanabashi \textit{et al.} [Particle Data Group],
Phys. Rev. D \textbf{98}, no.3, 030001 (2018)
doi:10.1103/PhysRevD.98.030001

\bibitem{Ezoe:2016mkp}
T.~Ezoe and A.~Hosaka,
Phys. Rev. D \textbf{94}, no.3, 034022 (2016)
doi:10.1103/PhysRevD.94.034022
[arXiv:1605.01203 [nucl-th]].

\bibitem{Ezoe:2017dnp}
T.~Ezoe and A.~Hosaka,
Phys. Rev. D \textbf{96}, no.5, 054002 (2017)
doi:10.1103/PhysRevD.96.054002
[arXiv:1703.01004 [hep-ph]].

\bibitem{Skyrme:1958vn}
T.~Skyrme,
Proc. Roy. Soc. Lond. A \textbf{A247}, 260-278 (1958)
doi:10.1098/rspa.1958.0183

\bibitem{Skyrme:1961vq}
T.~Skyrme,
Proc. Roy. Soc. Lond. A \textbf{A260}, 127-138 (1961)
doi:10.1098/rspa.1961.0018

\bibitem{Perring:1962vs}
J.~K.~Perring and T.~H.~R.~Skyrme,
Nucl.\ Phys.\  {\bf 31}, 550 (1962).
doi:10.1016/0029-5582(62)90774-5

\bibitem{Skyrme:1962vh}
T.~H.~R.~Skyrme,
Nucl.\ Phys.\  {\bf 31}, 556 (1962).
doi:10.1016/0029-5582(62)90775-7

\bibitem{Witten:1983tw}
E.~Witten,
Nucl. Phys. B \textbf{223}, 422-432 (1983)
doi:10.1016/0550-3213(83)90063-9

\bibitem{Witten:1983tx}
E.~Witten,
Nucl. Phys. B \textbf{223}, 433-444 (1983)
doi:10.1016/0550-3213(83)90064-0

\bibitem{Adkins:1983ya} 
G.~S.~Adkins, C.~R.~Nappi and E.~Witten,
Nucl.\ Phys.\ B {\bf 228}, 552 (1983).
doi:10.1016/0550-3213(83)90559-X

\bibitem{Zahed:1986qz}
I.~Zahed and G.~Brown,
Phys. Rept. \textbf{142}, 1-102 (1986)
doi:10.1016/0370-1573(86)90142-0

\bibitem{Callan:1985hy} 
C.~G.~Callan, Jr. and I.~R.~Klebanov,
Nucl.\ Phys.\ B {\bf 262}, 365 (1985).
doi:10.1016/0550-3213(85)90292-5

\bibitem{Callan:1987xt}
C.~G.~Callan, Jr., K.~Hornbostel and I.~R.~Klebanov,
Phys.\ Lett.\ B {\bf 202}, 269 (1988).
doi:10.1016/0370-2693(88)90022-6

\bibitem{Ring:1980aa}
P.~Ring and P.~Shuck,
``The Nuclear Many Body Problem'',
Springer, New York (1980).

\bibitem{Hosaka:2001ux}
A.~Hosaka and H.~Toki,
``Quarks, baryons and chiral symmetry,''
World Scientific (2001).
doi: https://doi.org/10.1142/4708

\end{thebibliography}
\end{document}